\documentclass[%
reprint,
longbibliography,
 amsmath,amssymb,
 aps,
]{revtex4-1}

\usepackage{graphicx}
\usepackage{dcolumn}
\usepackage{bm}
\usepackage{amsmath}
\usepackage{color}

\begin{document}


\title{Two-photon optical frequency reference with active ac Stark shift cancellation}

\author{V.~Gerginov}
\email{Vladislav.Gerginov@nist.gov.}
\affiliation{Department of Physics, University of Colorado, Boulder, CO 80309, USA}
\affiliation{Time and Frequency Division, National Institute of Standards and Technology, 325 Broadway, Boulder, CO 80305, USA}%

\author{K.~Beloy}

\affiliation{Time and Frequency Division, National Institute of Standards and Technology, 325 Broadway, Boulder, CO 80305, USA}%

\date{\today}

\newcommand{\ground}{\ensuremath{5s\,^{2}\!{S}_{1/2}}}
\newcommand{\excited}{\ensuremath{5d\,^{2}\!{D}_{5/2}}}
\newcommand{\coupled}{\ensuremath{5p\,^{2}\!{P}_{3/2}}}
\newcommand{\otherp}{\ensuremath{5p\,^{2}\!{P}_{1/2}}}

\begin{abstract}
An optical reference based on a two-photon optical transition with ac Stark shift cancellation is proposed. The reference uses two interrogating laser fields at different frequencies. Compared to conventional optical two-photon references, the new approach offers the possibility for improved short-term stability resulting from a higher signal-to-noise and improved long-term stability due to active ac Stark shift cancellation. We demonstrate the ac Stark shift cancellation method on the  $5s$\,$^2S_{1/2} \rightarrow 5d$\,$^2D_{5/2}$ two-photon transition in $^{87}$Rb.

\end{abstract}

\pacs{Valid PACS appear here}
\maketitle


\section{\label{sec:intro} Introduction}

The rubidium two-photon $5s$\,$^2S_{1/2} \rightarrow 5d$\,$^2D_{5/2}$ optical transition at 778\,nm has been recommended as a secondary representation of the second \cite{1984} and has been investigated as an optical frequency reference \cite{ZhuStandridge1997,PoulinLatrasseTouahriEtAl2002,DucosHonthaasAcef2002,EdwardsBarwoodMargolisEtAl2005, PerrellaLightAnstieEtAl2013, MartinPhelpsLemkeEtAl2018}. The reference is of particular interest because of the relative simplicity of the setup, high atomic $Q$-factor, and the possibility of using telecommunication components at 1560\,nm for both excitation and frequency division down to the microwave domain by optical frequency combs. The largest source of systematic frequency shift of the transition is the ac Stark shift (light shift), which also leads to frequency instability through changes in the interrogating laser intensity experienced by the atoms. The short-term stability of existing two-photon rubidium references has been limited to above $10 ^{-13}/ \sqrt{\tau}$ (with $\tau$ the integration time in seconds) due to the low detection efficiency of the employed fluorescence detection of the transition. Also, intensity and atomic temperature instability put upper limits on the interrogating laser intensity and atomic vapor density in order to reach the required long-term performance of the reference. 

In contrast to the traditional single-frequency (or single-color) interrogation, the two-photon transitions can also be driven by photons at two different optical frequencies (or two-color) \cite{BjorkholmLiao1974} that are tuned closer to a dipole-allowed transition. The clear advantage is the many orders of magnitude enhancement of the transition rate, at the expense of some transition broadening due to the residual Doppler shift \cite{BjorkholmLiao1976}. The technique has been used for sum frequency stabilization \cite{AkulshinHallIvannikovEtAl2011}, and in an absolute frequency reference \cite{PerrellaLightAnstieEtAl2013} via a self-referenced frequency comb. The enhancement in transition rate is accompanied with increased ac Stark shift due to the stronger dipole-allowed transition coupling. 

We can take advantage of the additional degrees of freedom offered by two-color excitation to minimize this systematic effect. First, as shown below, for a fixed transition rate the ac Stark shift is reduced. Second, as has been pointed out, the use of two different frequencies allows two-photon spectroscopy with zero differential ac Stark shift \cite{LETOKHOV1987}. The concept is similar to the magic wavelength in optical lattice clocks \cite{YeKimbleKatori2008}. By selecting an appropriate frequency detuning from the intermediate atomic state(s) coupled by allowed dipole transitions to the two-photon clock states, as well as the intensity ratios of the two optical fields, the differential ac Stark shift of the clock states can be zeroed. 

This proof-of-concept work demonstrates that the ac Stark shift of the  $5s$\,$^2S_{1/2} \rightarrow 5d$\,$^2D_{5/2}$ two-photon Rb optical reference can be actively canceled by the use of two-color spectroscopy. The detection of the atomic transition is more efficient because the transition branching ratios do not play a role in contrast to fluorescence detection, the efficiency of a Si photodetector is higher than that of a photomultiplier, and the reduced detuning from the intermediate state allows the laser beam waist (and correspondingly active interaction volume) to be increased to achieve the same transition rate as in the single-color case. The results pave the way to both improved short- and long-term performance of the optical frequency reference. 


\section{\label{theory}Transition rate and ac Stark shift}

For the present section, we label the participating states as
$\left|g\right\rangle\equiv\left|\ground\right\rangle$,
$\left|e\right\rangle\equiv\left|\excited\right\rangle$,
$\left|k\right\rangle\equiv\left|\coupled\right\rangle$,
and $\left|l\right\rangle\equiv\left|\otherp\right\rangle$. Letting $\left|i\right\rangle$ and $\left|j\right\rangle$ be generic, we further introduce the transition frequencies $\omega_{ij}\equiv\left(E_i-E_j\right)/\hbar$ and $\mathcal{D}_{ij}\equiv\left\langle i||\mathbf{D}||j\right\rangle$, where $E_i$ is the energy of state $\left|i\right\rangle$, $\hbar$ is the reduced Planck constant, and $\mathbf{D}$ is the electric dipole operator. The reduced matrix elements $\mathcal{D}_{ij}$ are taken as real. The frequencies $\omega_{ij}$ are given in \cite{BARWOODPROWLEY1991, YeSwartzJungnerEtAl1996, MaricMcFerranLuiten2008}, while theoretical values for the $\mathcal{D}_{ij}$ are given in \cite{SafronovaWilliamsClark2004}.

The clock transition $\left|g\right\rangle\rightarrow\left|e\right\rangle$ is driven by two counter-propagating laser beams with lin\,$\perp$\,lin polarization. The laser beams have frequencies $\omega_1$ and $\omega_2$, with respective intensities $I_1$ and $I_2$ at the atoms. The frequencies $\omega_1$ and $\omega_2$ are well-detuned from electric dipole transition frequencies $\omega_{kg}$, $\omega_{lg}$, and $\omega_{ek}$ relative to natural linewidths and hyperfine splittings. Hyperfine levels of the clock states, specified by $F_g$ and $F_e$, are resolved by the two-photon transition. A bias magnetic field (1\,$\mu$T is chosen exprimentally) aligned along the laser beam axis is sufficiently large to define the quantization axis, though accompanying Zeeman splittings are unresolved. The initial ground state population is evenly distributed among the magnetic substates. From second-order time-dependent perturbation theory, the transition rate reads
\begin{equation}
\begin{aligned}
w_{g\rightarrow e}={}&
I_1I_2\frac{A\pi\mathcal{D}_{ek}^2\mathcal{D}_{kg}^2}{2\hbar^4\epsilon_0^2c^2}
\left(\frac{1}{\omega_{kg}-\omega_1}+\frac{1}{\omega_{kg}-\omega_2}\right)^2
\\&\times
\delta\left(\omega_1+\omega_2-\omega_{eg}\right),
\end{aligned}
\label{Eq:transrate}
\end{equation}
where $\epsilon_0$ is the permittivity of free space and $c$ is the speed of light. The Dirac $\delta$-function appearing here is a proxy for a lineshape function of finite bandwidth. Finally, the factor $A$ depends on the initial and final hyperfine levels, with values specified in Table~\ref{Tab:angularfactors}.

\begin{table}
\caption{Factors $A$ and $B$, derived from basic principles of angular momentum theory, for the various transitions $F_g\rightarrow F_e$, where $F_g$ and $F_e$ label the hyperfine level of the ground and excited clock states, respectively.}
\label{Tab:angularfactors}
\begin{ruledtabular}
\begin{tabular}{cccccc}
$F_g\rightarrow F_e$	& $A$	& $B$	& $F_g\rightarrow F_e$	& $A$	& $B$	\\
\hline
\vspace{-2mm}\\
$1\rightarrow 1$		&3/800	&7/50	& $2\rightarrow 1$		&1/4000	& $-7/50$	\\
$1\rightarrow 2$		&7/1440	&1/14	& $2\rightarrow 2$		&1/800	& $-3/98$	\\
$1\rightarrow 3$		&7/1800	&33/175	& $2\rightarrow 3$		&7/2000	& $33/700$	\\
						& 		&		& $2\rightarrow 4$		&3/400	& $55/196$	\\
\end{tabular}
\end{ruledtabular}
\end{table}

While the two lasers drive the clock transition, they also induce ac Stark shifts to the clock levels,
\begin{equation}
\begin{gathered}
\delta E_g=
-(I_1/2\epsilon_0c)\alpha_g\left(\omega_1\right)
-(I_2/2\epsilon_0c)\alpha_g\left(\omega_2\right),
\\
\delta E_e=
-(I_1/2\epsilon_0c)\alpha_e\left(\omega_1\right)
-(I_2/2\epsilon_0c)\alpha_e\left(\omega_2\right),
\end{gathered}
\label{Eq:dEboth}
\end{equation}
with AC polarizabilities given by
\begin{gather*}
\alpha_g\left(\omega\right)
=\frac{1}{3\hbar}\left(
\mathcal{D}_{kg}^2
\frac{\omega_{kg}}{\omega_{kg}^2-\omega^2}
+\mathcal{D}_{lg}^2
\frac{\omega_{lg}}{\omega_{lg}^2-\omega^2}
\right),
\\
\alpha_e\left(\omega\right)
=-\frac{1}{9\hbar}\left(1+B\right)\mathcal{D}_{ek}^2
\frac{\omega_{ek}}{\omega_{ek}^2-\omega^2}.
\end{gather*}
For the ground clock state, this corresponds to the conventional scalar polarizability. For the excited clock state, both scalar and tensor polarizabilities contribute, with the factor $B$ here accounting for the latter. As with $A$, the factor $B$ depends on the initial and final hyperfine levels, with values specified in Table~\ref{Tab:angularfactors}.

Introducing the quantity
\begin{equation*}
R\equiv-\frac
{\alpha_e\left(\omega_2\right)-\alpha_g\left(\omega_2\right)}
{\alpha_e\left(\omega_1\right)-\alpha_g\left(\omega_1\right)},
\end{equation*}
we see from Eq.~(\ref{Eq:dEboth}) that the clock frequency is unperturbed when operating with an intensity ratio $I_1/I_2=R$. For the case $\omega_1=\omega_2$, the clock shift is simply proportional to the sum of intensities, $I_1+I_2$, and no intensity ratio can null it. This is reflected in the negative result for $R$, namely $R=-1$. For $\omega_1\neq\omega_2$, on the other hand, $R$ can accommodate positive values, allowing the clock shift to be precisely nulled with the appropriate intensity ratio. To explore this further, we assume the resonance condition $\omega_1+\omega_2=\omega_{eg}$ and parametrize the laser frequencies with detuning $\Delta$ according to \begin{gather*} \omega_1=\omega_{kg}+\Delta, \\ \omega_2=\omega_{ek}-\Delta.
\end{gather*}
In the limit $\Delta\rightarrow0$, we obtain the result $R\rightarrow\left[\left(1+B\right)/3\right]\left(\mathcal{D}_{ek}/\mathcal{D}_{kg}\right)^2$, which is positive for each transition in Table~\ref{Tab:angularfactors}. Increasing $\Delta$ moves towards the case $\omega_1=\omega_2$ and negative values of $R$.
Taking transition frequencies $\omega_{ij}$ from Ref.~\cite{BARWOODPROWLEY1991, YeSwartzJungnerEtAl1996, BernardMadejSiemsenEtAl2000} and dipole matrix elements $\mathcal{D}_{ij}$ from Ref.~\cite{SafronovaWilliamsClark2004} and expanding to second order in $\Delta$, we find for the $F_g=2$, $F_e=4$ transition \begin{equation*} R=(0.0656)\left[1-\left(8.06\times10^{-3}\right)\Delta-\left(3.19\times10^{-6}\right)\Delta^2\right],
\end{equation*}
where $\Delta$ is in units of $2\pi\times\mathrm{GHz}$. The second order expansion here is an excellent approximation through the zero crossing at $\Delta/2\pi=118$~GHz. In practice, the choice $\Delta/2\pi=10$~GHz marks a good compromise; this is sufficiently well-detuned from the electric dipole transitions, while an intensity ratio of $I_1/I_2=0.0602$ can be used to drive the clock transition with zero ac Stark shift. For this operational condition, a 1\% single laser beam intensity change leads to a change in the ac Stark shift a factor of two lower than in the single-color case with equal beam intensities and identical transition rate (accounting only for transitions with a photon absorbed from each counter-propagating beam).

\section{\label{two}Two-photon Rb spectroscopy}

\begin{figure}
\includegraphics[width=\columnwidth]{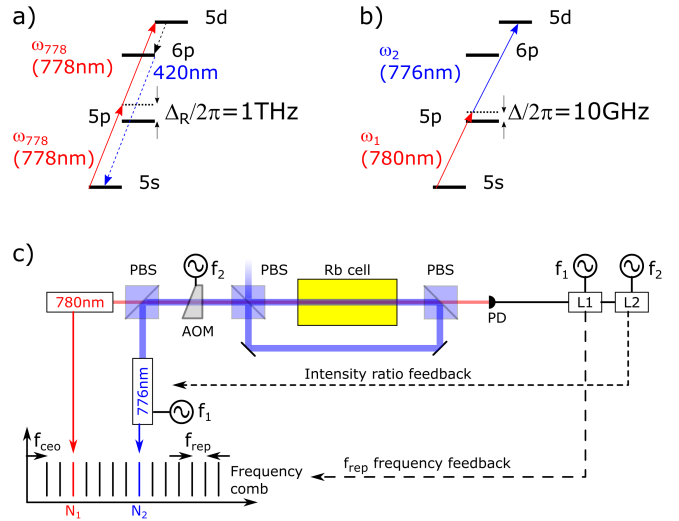}
\caption{\label{fig:Fig1} Simplified energy diagram of Rb two-photon transition and experimental setup. a) Conventional single-color case. b) Two-color case used in this work. c) Experimental setup for the proposed optical reference implementing ac Stark shift cancellation. PBS - polarizing beam splitter, BS - beam splitter, AOM - acousto-optic modulator, PD - photodetector, L1, L2 - lock-in amplifiers.}
\end{figure}

Figure~\ref{fig:Fig1} a) shows a simplified Rb energy diagram of the conventional two-photon transition interrogation scheme using a single-color light field. The light is blue-detuned from the closest $5p$\,$^2P_{3/2}$ state by $\Delta_R/2 \pi \approx $1\,THz. The $5d$\,$^2D_{5/2}$ state decays through the $6p$\,$^2P_{3/2}$ intermediate state, and the 420\,nm fluorescence from the $6p$\,$^2P_{3/2}$ state decay is detected (typically by a photomultiplier). The laser frequency is stabilized to the atomic transition, and its frequency can be used to reference an optical frequency comb by the relation $\omega_{2ph}=2 \omega_{778}=2 \pi \times \left(2N_{778} \times f_{rep} +2 f_{CEO} \right)$, where $\omega_{2ph}$ is the two-photon transition frequency and $\omega_{778}$ the laser optical frequency (in radians/second), $N_{778}$ is the mode number of the comb mode locked to the laser frequency $\omega_{778}$, and $f_{rep}$ and $f_{CEO}$ are the repetition rate and the carrier-envelope offset frequency of the self-referenced comb. 

Figure~\ref{fig:Fig1} b) shows the two-color two-photon transition interrogation scheme. Two light fields at wavelengths of 780\,nm and 776\,nm excite the transition when the sum of their optical frequencies $\omega_1$ and $\omega_2$ matches the two-photon frequency $\omega_{2ph}$. The 780\,nm light is blue-detuned ($\Delta/2 \pi=10$\,GHz) from the $5s$\,$^2S_{1/2} \rightarrow 5p$\,$^2P_{3/2}$ transition at 780\,nm for reasons we discuss below. The sum of the optical frequencies is stabilized to the two-photon atomic transition. If both light fields are locked to respective optical components of a self-referenced optical frequency comb \cite{PerrellaLightAnstieEtAl2013}, the single degree of freedom  - the comb repetition rate - can be referenced to the atomic transition frequency by the relation $\omega_{2ph}=\omega_{1}+\omega_{2}=2 \pi \times \left((N_{1}+N_{2}) \times f_{rep} +2 f_{CEO}\right)$, with $N_{1}$ and $N_{2}$ the two frequency comb components to which $\omega_{1}$ and $\omega_{2}$ are locked.

Fig.~\ref{fig:Fig1} c) outlines the scheme of the proposed optical frequency reference implementing active ac Stark shift cancellation. Two lasers are locked to the respective teeth $N_1$ and $N_2$ of a self-referenced frequency comb. The optical frequency of the 776\,nm laser is frequency-modulated (FM) at a frequency $f_1$. The output beams of both lasers are collimated, linearly polarized, and overlapped to propagate in the same direction with orthogonal polarizations. The two beams pass a common modulator (shown in our example is an acousto-optic modulator) performing same-depth intensity modulation of both laser beams at a frequency $f_2$. The two intensity-modulated beams are spatially separated using a polarizing beamsplitter, and recombined in a Rb vapor cell in a counter-propagating configuration. The 780\,nm laser beam is detected, and demodulated at the frequency $f_1$ (a modulation transfer spectroscopy \cite{Shirley1982}) with the lock-in amplifier L1, providing a feedback signal for locking the repetition rate $f_{rep}$ of the self-referenced frequency comb to the atomic transition. A second demodulation of the L1 output signal feedback signal is performed with the lock-in amplifier L2 at the intensity modulation frequency $f_2$. The output of the lock-in amplifier L2 provides an intensity ratio feedback signal derived from the atomic resonance response to the intensity modulation. This feedback signal can be used to keep the intensity ratio at the value that cancels the differential ac Stark shift experienced by the atoms by changing the output power of one of the lasers (shown in our example is the 776\,nm laser).

The use of the two-color scheme for two-photon Rb spectroscopy has some drawbacks. The use of two lasers requires control of the laser frequency difference when the sum frequency is stabilized to the atomic transition. As the reference is intended to operate in combination with a frequency comb (see Fig.\,\ref{fig:Fig1} c)), the frequency difference can be controlled by referencing the lasers to the comb. The effective atomic transition linewidth is increased due to the contributions from residual Doppler broadening and both laser linewidths (which can be made negligible with existing telecom-based lasers). The residual Doppler broadening also reduces the amount of atoms interacting with the light by a factor of three due to velocity class selection. At the same time, the increased detection efficiency and interaction volume compensate for these two effects. The detection efficiency is $\sim10$ times higher since the branching ratios do not play a role. The transition rate is proportional to $1/\Delta^2$ and is increased by a factor of $\sim 5 \times 10^3$ in the two-color case, allowing a reduction of the laser intensities and correspondingly larger interaction volume under the condition of same total laser power and transition rate. The intensity modulation depth required for the ac Stark shift cancellation must be the same for the two laser beams of different frequencies, which can be accomplished by using the same modulator. Finally, the signal detection has a contribution from the intensity noise and photon shot noise of the 780\,nm laser beam, and an intensity noise suppression might be required. 

\section{\label{exp}Experiment}

\begin{figure}
\includegraphics[width=\columnwidth]{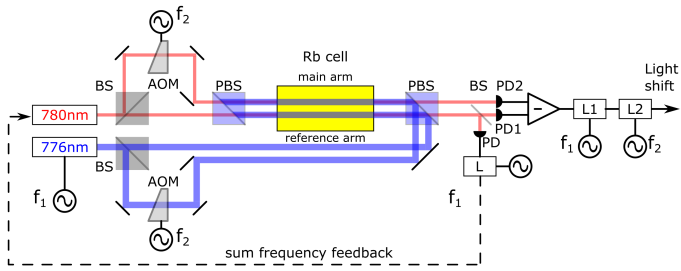}
\caption{\label{fig:Fig2} Experimental setup for ac Stark shift cancellation study. The 776\,nm and 780\,nm laser beams propagate independently and have lin $\perp$ lin polarizations. The laser intensities can be modulated independently with AOMs. The reference and main arms share the same vapor cell to suppress additional common-mode noise and shifts. The lock-in amplifier L is used for sum frequency locking; lock-in amplifiers L1 and L2 are used to study the ac Stark shift in the main spectroscopy arm.}
\end{figure}

The proof-of-concept experimental setup for ac Stark shift cancellation is shown in Fig.~\ref{fig:Fig2}. The optical fields at 776\,nm and 780\,nm wavelengths are created by commercial (Littrow configuration) and home-made (Littman configuration) external-cavity diode lasers, each with typical linewidth on the order of 100\,kHz. Because of the significant laser frequency noise, a differential measurement is made between a main spectroscopy arm used to study the ac Stark shifts, and a reference spectroscopy arm used to lock the sum laser frequency to the two-photon transition and to suppress the common-mode noise between the main and the reference arms. The two independent optical arms with overlapping counter-propagating orthogonally-polarized laser beams are created in a single Rb cell. The 25\,mm diameter, 25\,mm length cell is filled with isotopically-pure $^{87}$Rb and heated to $100\,^\circ$C. The cell is placed in a single-layer magnetic shield. The small magnetic bias field ($\sim$\,1\,$\mu$T) is applied along the direction of the laser beams.

To demonstrate the ac Stark shift of the two-photon transition caused by the lasers, the laser beams propagate and are intensity-modulated independently, in contrast to the proposed scheme of Fig.\,\ref{fig:Fig1} c). The first laser can be tuned by $\sim 20$\,GHz around 776.000(1)\,nm and provides 25\,mW of light at the output of a polarization-maintaining (PM) fiber. The laser current is modulated at frequency $f_1=16.5$\,kHz with a modulation depth of 1.5\,MHz. The second laser can be tuned by $\sim 2$\,GHz around 780.221(1)\,nm and provides 1.5\,mW of light at the output of another PM fiber. The frequency modulation of the first laser causes a corresponding modulation of the transition rate, which in turn leads to a modulation of the transmitted light from the second laser that is used for the detection of the atomic transition \cite{Shirley1982}.

The reference arm is used for sum laser frequency locking using modulation-transfer spectroscopy. The 780\,nm light is detected with the photodetector PD and demodulated with the lock-in amplifier L. The L output is used to lock the frequency of the 780\,nm laser to that of the free-running 776\,nm laser so that the sum laser frequency equals the two-photon transition frequency. Proportional and integral feedbacks are used for fast (laser diode injection current) and slow (external cavity piezo actuator) frequency control. 

The main arm is used to introduce and detect ac Stark shifts. Two acousto-optic modulators (AOMs) are used in zero-order to modulate independently the intensity of each laser beam at a frequency $f_2=200$\,Hz. The AOM modulations are in phase. The lock-in amplifier L1 demodulates the 780\,nm light frequency modulation at the frequency $f_1$. The lock-in amplifier L1 uses the same reference and settings as L. A differential frequency shift detection between both arms is based on a balanced photodetector which subtracts parts of the 780\,nm signals from each arm. Through the ac Stark effect, an AOM-modulated laser intensity causes a corresponding modulation of the transition frequency in the main arm and at the L1 output. The lock-in amplifier L2 detects the modulation due to the intensity modulation(s) present at the L1 output at the frequency $f_2$. The noise from common-mode sources (such as laser frequency and amplitude noise) in the two arms is suppressed. The output of L2 provides an error signal which could be used to control the intensity ratio of the 776\,nm to 780\,nm beams. The amplitude of the error signal depends on the intensity and the modulation depth of each laser beam. For equal modulation depths, the error signal is proportional to the deviation of the intensity ratio from the value corresponding to a zero ac Stark shift.

\begin{figure}
\includegraphics[width=0.65\columnwidth]{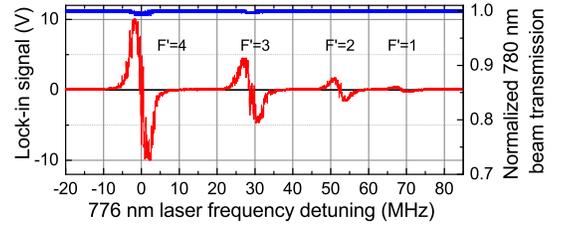}
\caption{\label{fig:spectrum} Two-photon $5s$\,$^2S_{1/2} (F=2) -5d$\,$^2D_{5/2}$ spectrum in $^{87}$Rb using two-color spectroscopy. Dispersive signal (red) - left scale, transmission signal (blue) - right scale.}
\end{figure}

A two-color two-photon $^{87}$Rb spectrum  measured with the photodetector PD and the lock-in amplifier L, is shown in Fig.~\ref{fig:spectrum}. The laser powers were $0.7$\,mW (at 780\,nm) and $14$\,mW (at 776nm), with beam waist of 2\,mm. The detuning from the $6p$\,$^2P_{3/2}$ state was $\Delta /2 \pi=10$\,GHz. The laser frequency noise contribution is clearly visible on the $\sim 3.8$\,MHz wide transitions, broadened by the residual Doppler effect resulting from the frequency difference of the two counter-propagating beams \cite{BjorkholmLiao1976}. The lock-in amplifier L had a time constant of 300\,$\mu$s. With the 776\,nm laser detuned from resonance, measurements of the power spectral density at the L output resulted in 440\,$\mu \text{V$_{\text{rms}}$}$/Hz$^{1/2}$, a factor of 2.2 above the electronics noise (measured with the light blocked on the PD). This noise level corresponds to a fractional frequency instability of the sum frequency of $7.7 \times 10^{-14}/\sqrt{\tau}$, obtained using a frequency discriminator slope of $134$\,kHz/V determined by the slope of the $5s$\,$^2S_{1/2} (F_g=2) -5d$\,$^2D_{5/2} (F_e=4)$ spectral component.

With the sum laser frequency locked to the transition via the reference arm, the ac Stark shift was investigated with the main arm of the spectroscopy setup. The intensity ratio was set close to the theoretically determined value for zero ac Stark shift by measuring the individual laser beam powers and assuming the same beam intensity profiles. A 5\,\% sinusoidal amplitude modulation (AM) at $f_2=200$\,Hz was introduced to each laser beam in the main arm with independent AOMs. The modulation depths were equalized to within 10\,\% by controlling the amplitude of the RF signals delivered to the AOMs. The two AM modulations were phase-synchronized during the experiment, and the phase of the lock-in amplifier L2 with respect to the reference frequency $f_2$ was set to zero. The time constant of the lock-in amplifier L2 was 30\,s. The output of the lock-in amplifier L2 was used as a measure of the modulation-induced ac Stark shift (error signal, expressed as fractional frequency deviation).
The conversion between voltage at the output of lock-in amplifier L2 and fractional frequency deviation was performed by using the slope of the frequency discriminator signal (example shown in Fig.~\ref{fig:spectrum}), and the sensitivity settings of the lock-in amplifier L2.

\begin{figure}
\includegraphics[width=0.9\columnwidth]{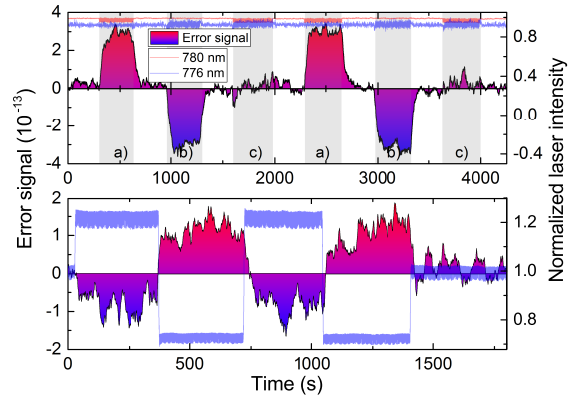}
\caption{\label{fig:Fig4} Error signal (ac Stark shift expressed as fractional frequency deviation) and normalized laser intensities (right scale) as a function of time. Top plot shows the effect of laser intensity modulation of a) 780\,nm beam; b) 776\,nm beam; c) both beams on the error signal. Bottom plot shows the effect of the total intensity change of the 776\,nm beam on the error signal.}
\end{figure}

The effect of the intensity modulation causing ac Stark shift is illustrated in Fig.~\ref{fig:Fig4}, top plot. Three cases - a) to c) - can be distinguished. In case a), the 780\,nm beam is intensity-modulated. The error signal is positive - the transition frequency decreases with increased laser intensity. In case b), the 776\,nm beam is intensity-modulated. The error signal is negative, corresponding to an ac Stark shift of the transition which has an opposite sign compared to the shift in case a). In case c), both beams are intensity-modulated. The error signal is close to zero, since the ac Stark shifts caused by each light fields have equal magnitude (determined by the equal modulation depths and the chosen intensity ratio) and cancel, leaving the two-photon transition frequency undisturbed by the intensity modulations.

The effect of intensity ratio change is illustrated in Fig.~\ref{fig:Fig4}, bottom plot. Intensity modulation with the same depth was applied to both 780\,nm and 776\,nm beams. The power of the 776\,nm beam was varied (changing the intensity ratio as well as the absolute intensity modulation depth). The intensity ratio was changed within $50$\,\% around the zero ac Stark shift value. The error signal is negative when the intensity ratio of the 776\,nm and the 780\,nm beams is above the zero shift value, positive when the ratio is below the zero shift value, and can be used to actively stabilize the ratio to the zero ac Stark shift value.

The ac Stark shift signals shown in Fig.~\ref{fig:Fig4} are at the $10^{-13}$ level, showing signal to noise ratios of less than 5. The signal to noise ratio is mostly determined by residual laser frequency fluctuations, laser intensity fluctuations, and lock-in amplifier L2 time constant of 30\,s. We expect the residual ac Stark shift to be below the $10^{-14}$ level when narrower-linewidth lasers, reduced laser intensity noise, and longer time constants are used. We envision time constants longer than 100\,s as the ac Stark shift cancellation is intended to improve the long-term stability of the system.

Cancellation of the ac Stark shift has already been proposed for microwave frequency references \cite{HashimotoOhtsu1990, ShahGerginovSchwindtEtAl2006, McGuyerJauHapper2009}. In certain cases, the cancellation is performed without additional modulation apart from the one required for frequency locking \cite{HashimotoOhtsu1990, McGuyerJauHapper2009}. The techniques \cite{HashimotoOhtsu1990, McGuyerJauHapper2009} rely on significant atomic absorption to cause significant amplitude modulation, and on the atomic transition symmetry in order to extract information about the ac Stark shift and cancel it. In the experiment described in this work, frequency modulation of the 776\,nm light is used to lock the sum laser frequency to the atomic transition, and the expected atomic absorption will be above 1\%. It might be possible to make use of the resulting light amplitude modulation (AM) of the light fields resulting from the FM-to-AM conversion in the presence of frequency-dependent atom absorption, and apply the ac Stark shift cancellation method proposed here without an additional intensity modulation.

\section{\label{comparison} Comparison between single-color and two-color two-photon spectroscopy.}

A parameter comparison between a single-color and two-color two-photon optical frequency references is done using the results from Section \ref{theory} and is given in Table~\ref{tab:table1}. We consider a detuning $\Delta/2 \pi =10$\,GHz for the two-color case, and impose the condition of equal transition rates for the single-color and the two-color schemes. The Rb cell is assumed $L=100$\,mm long, at a temperature of 90\,$^{\circ}$C, corresponding to a pressure of 31\,mPa and a Rb vapor density of $3.1 \times 10^{12}$/cm$^3$. The natural linewidth of the two-photon transition is taken from  Ref.\,\cite{ShengPerezGalvanOrozco2008}. The stability at 1 second $\sigma_y\left(\text{1\,s}\right)$ is calculated in fractional frequency units according to the formula
\begin{equation}
	\sigma_y\left(\text{1s}\right)=\frac{\delta \omega}{\omega_{\text{2ph}}}\frac{1}{\text{SNR}}
\end{equation}
where $\delta \omega$ is the effective transition linewidth and SNR is the signal-to-noise ratio (given by the square root of the number of detected atoms per second in the quantum projection noise (QPN)-limited case).

For the single-color setup (Fig.~\ref{fig:Fig1} a), the laser beams are assumed focused to a waist $w_0=0.4$\,mm and have $10$\,mW power per beam, corresponding to a Rayleigh length $z_R=\pi w_0^2/\lambda_{778}=0.7$\,m. It is assumed that the beam waist does not change significantly over the cell length, and the active volume is calculated as $\pi w_0^2 L / 4$. With a thermal velocity of 170\,m/s, the transit time broadening is 130\,kHz. Fluorescence detection is used, as detection of a small signal on a large background is challenging. The efficiency of the transition detection for this setup is reduced by the branching ratio of the excited state, with only 35\,\% of the excited state decaying to $6p$\,$^2P_{3/2}$ state, and 31\,\% of the atoms in this state emitting 420\,nm photons \cite{Heavens1961}. The detection efficiency is given as the product of the branching ratios, the fluorescence light collection efficiency of the optical system (typically 20\%), and the PMT quantum efficiency (typical 30\,\% at 420\,nm). The additional signal loss due to resonant 420\,nm fluorescence re-absorption is neglected. The ac Stark shift for 1\,\% single laser beam intensity change is calculated using the formulas from Section \ref{theory}.

For the two-color setup (Fig.~\ref{fig:Fig1} b)), the laser beams are assumed collimated to $d=2$\,mm diameter, with powers $P_{780}=1.2$\,mW and $P_{776}=19.2$\,mW, and intensity ratio as required for the ac Stark shift cancellation. The active volume is calculated as $\pi d^2 L / 4$. The transit time broadening is reduced to 30\,kHz due to the larger beam diameter. Because of the different wavelengths of the two laser beams, the residual Doppler broadening of the two-photon transition is $\left(\omega_{2}-\omega_{1}\right)\Delta f_{2ph}/\omega_{\text{2ph}} =2.76$\,MHz, with $\Delta f_{2ph}$ the Doppler width of the two-photon transition \cite{BjorkholmLiao1976}. Because of the increased interaction volume, the signal can be efficiently detected in absorption \cite{LevensonEesley1979}. The absorption of the 780\,nm light is preferable as the  ratio of photon scattering rate to the beam photon flux is larger than that at 776\,nm (as well as at 778\,nm in the single-color case). There is no re-absorption of the detected 780\,nm light except due to resonant two-photon excitation (in contrast to the fluorescence at 420\,nm). In this case the branching ratios do not play a role, and the light collection efficiency is 100\,\%. A typical Si photodiode quantum efficiency is 65\,\% at 780\,nm. The number of detected atoms per second is reduced by a factor of three due to the residual Doppler effect causing excitation of a specific velocity class. The stability at 1 second is more than seven times higher than in the single-color case, and the ac Stark shift for 1\,\% single laser beam intensity change is two times lower than the value in the single-color case.

\begin{table}
\caption{\label{tab:table1} Two-photon Rb optical reference parameters. Second column - single-color case. Third column - two-color case. $\sigma_B\left(\text{LS}\right)$ is the fractional frequency uncertainty caused by 1\,\% variation of single-beam intensity.}
\begin{ruledtabular}

\begin{tabular}{lll}
Parameter																			& Single-color														&  Two-color													\\
\hline
Wavelength (nm)																& 778 / 778																&  776 / 780															\\
Frequency $\omega_{\text{2ph}}/2 \pi \text{ (kHz)}$                                       & \multicolumn{2}{c}{$2\times$385\,285\,142\,375} \\
Laser power (mW)															& 10.0 / 10.0															& 19.2 / 1.2															\\
Beam waist (mm)																& 0.4																			& 2																				\\
Active volume (mm$^3$)												& 13																			& 314																			\\
Virt. state det.  $\Delta_R/2 \pi \text{(GHz)}$ 			& 1054														& 10																			\\
Nat. linewidth (MHz)													& 0.667																		& 0.667																		\\
Res. Doppler width (MHz)											& n/a																			& 2.79																		\\
Transit broadening (MHz)											&	0.128																		& 0.027																		\\
Eff. linewidth $\delta \omega / 2 \pi \text{ (MHz)}$			& 0.795																		& 3.48												\\
Cell temperature $(^{\circ}\text{C})$					& 90																			& 90																			\\
Branching ratio																&	$0.35 \times 0.31 $											&	n/a																			\\
Light collection efficiency										& $0.2$																		&	n/a																			\\
Quantum efficiency														&	$0.3$ (PMT)															& $0.65$	(Si) 														\\
SNR, $\times 10^6$														&	$0.5 $																	&	$28 $																		\\				
$\sigma_y\left(\text{1\,s}\right)$, $\times 10^{-15}$									&	2.2							& 0.3					 	       				 						\\
$\sigma_B\left(\text{LS}\right)$ (1\,\%), $\times 10^{-13}$							& 1.8 					& 0.87 																		\\

\end{tabular}
\end{ruledtabular}
\end{table}

\section{Conclusions}

The results from Section \ref{comparison} show that for the same transition rate, the stability $\sigma_A\left(\text{1\,s}\right)$ of the two-color scheme is seven times better than the single-color one due to the higher detection efficiency. The absorption measurement eliminates the need for a photomultiplier typically used in fluorescence detection. The 1\,\% intensity change ac Stark shift systematic uncertainty $\sigma_B\left(\text{LS}\right)$, the largest source of systematic uncertainty in this type of reference, is a factor of two lower for the two-color reference compared to the single-color one. This uncertainty can be completely removed with the demonstrated ac Stark shift cancellation technique, making the reference more accurate. The two-laser scheme allows the use of modulation transfer spectroscopy. The use of collimated beams simplifies the experimental setup, as no build-up cavity or focusing are required, opening the possibility for a multi-pass cell arrangement that would improve the SNR and the reference stability. A frequency reference is characterized by its stability and accuracy, and the two-color scheme offers improvement of both.

In summary, the two-color scheme described in this work offers the possibility of building a rubidium frequency reference based on telecom components with the option of controlling the largest systematic uncertainty contributions to levels below $10^{-13}/\sqrt{\tau}$. Such a device would fulfill the growing need for optical frequency references that outperform their commercial microwave counterparts both short- and long-term, and can operate with relaxed environmental control requirements compared to optical references based on cold atoms.

\section{Acknowledgements}

The authors thank James Bergquist for technical help, and Elizabeth Donley and Richard Fox for their helpful comments on the manuscript. This work is a contribution of NIST, an agency of the US government, and is not subject to copyright in the US. 

\section{References}


%

\end{document}